\documentclass{ws-procs9x6-cpt19}
\begin{document}

\newcommand{\refeq}[1]{(\ref{#1})}
\def\etal {{\it et al.}}

\title{Mining the Data Tables for Lorentz and CPT Violation}

\author{N.E.\ Russell}

\address{Physics Department, Northern Michigan University,\\
Marquette, Michigan 49855, USA}

\begin{abstract}
In this conference proceedings,
some comments on the present status and recent growth 
of efforts to find Lorentz and CPT violation are given 
by extracting metrics from the annually updated 
{\it Data Tables for Lorentz and CPT Violation.} 
They reveal that tests 
span all the sectors of particle and gravitational physics,
and have shown remarkable and consistent growth.
Through numerous innovations and refinements in experiments,
a large body of data has been amassed
with ever-increasing precisions.
The Tables are available through the Cornell preprint archive at arXiv:0801.0287.

\end{abstract}

\bodymatter

\phantom{}
\vskip10pt
\noindent
The first edition of the 
{\it Data Tables for Lorentz and CPT Violation}
appeared in the Proceedings of the Fourth Meeting on Lorentz and CPT Symmetry
and on the Cornell preprint server arXiv.org, 
where it has been updated each year.\cite{datatables}
A July 2010 edition appeared in Reviews of Modern Physics.\cite{rmp2011}
The purpose of the Tables is to maintain a complete list 
of measurements of coefficients for Lorentz and CPT violation,
to summarize the maximal attained sensitivities in high-activity sectors,
and to provide information about properties and definitions 
relevant to the study of these fundamental symmetries.

With the Tables being in existence for more than ten years,
and on the occasion of this Eighth Meeting on Lorentz and CPT Symmetry 
being held more than 20 years since the first such meeting in 1998,
it is a good time to take a retrospective look at the 
extraordinary theoretical and experimental efforts 
that have been made by hundreds of researchers
since the 1990s to find evidence of Lorentz violation in nature.

\begin{figure}
\begin{center}
\includegraphics[width=180pt]{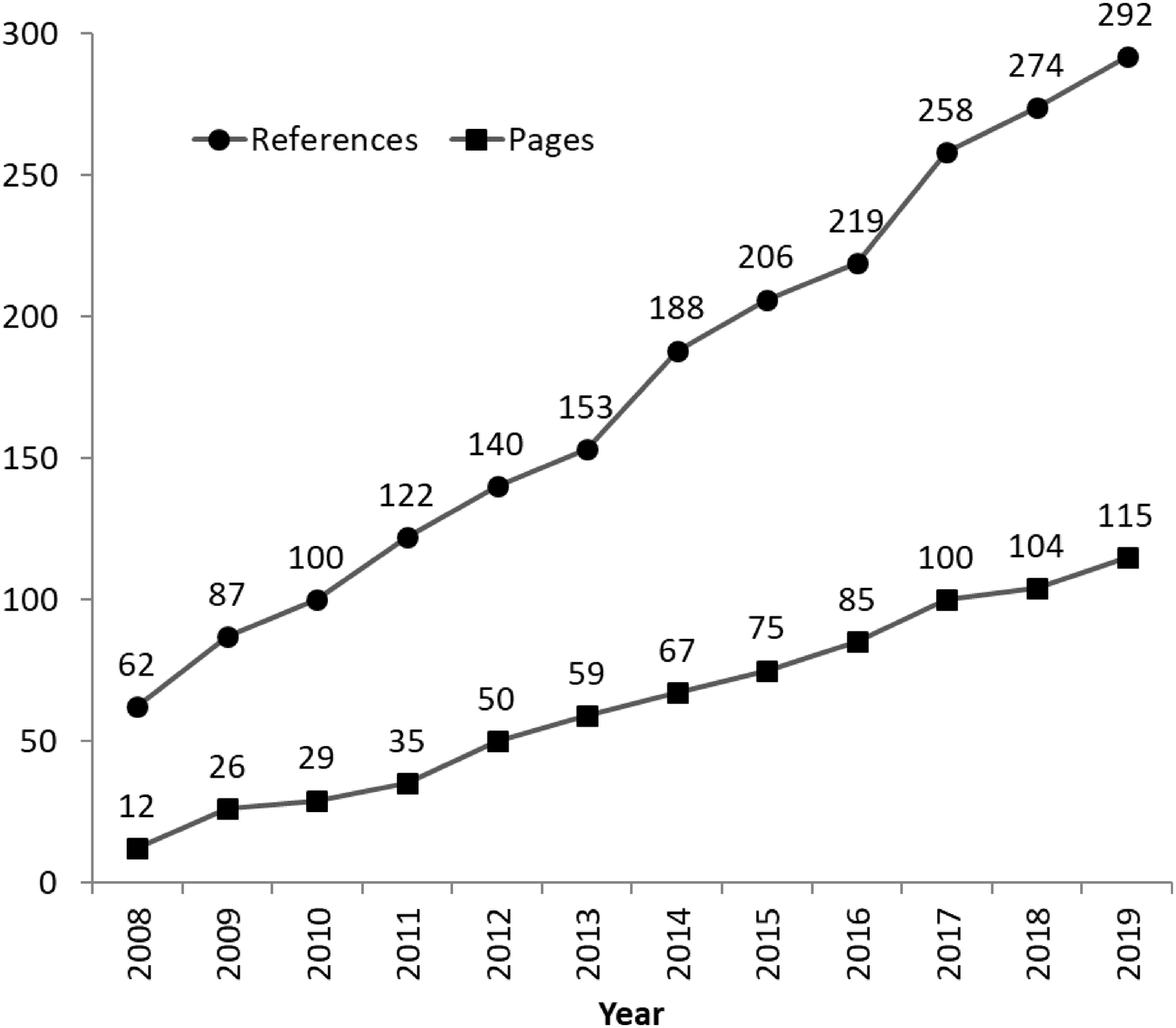}
\end{center}
\caption{The number of bibliographic references and the page count 
for the twelve editions of the {\it Data Tables for Lorentz and CPT Violation} 
up to January 2019.
}
\label{growth}
\end{figure}

The burgeoning number of publications placing limits on Lorentz and CPT symmetry
can be traced in the annual updates 
to the Tables.
The upper plot in Fig.\ \ref{growth}
shows the number of references in each edition of the Tables
growing five-fold from 62 in 2008 to 292 in 2019.
Another measure is the page count, 
which has grown about ten-fold from twelve to 115 
over the same period.

The Standard-Model Extension,
or SME,
categorizes Lorentz violation in 
the behavior of all known particles
and their interactions,\cite{dcak}
including gravitational ones.\cite{akgravity}
It is the result of considering effective field theory 
with Lorentz and CPT violation,
leading to a general framework 
incorporating known physics but also admitting violations 
of these symmetries.
The Tables list measurements of the 
coefficients controlling Lorentz- and CPT-breaking operators
of all mass dimensions in the framework.

\begin{figure}
\begin{center}
\includegraphics[width=170pt]{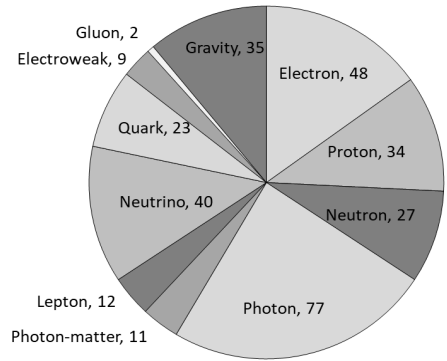}
\end{center}
\caption{The number of Data Table references presenting limits 
in each of the sectors of the Standard-Model Extension.}
\label{sectors}
\end{figure}

Figure \ref{sectors} indicates the distribution of efforts 
across the broad scope of the field. 
The size of each slice of the pie chart 
is set by the number of publications 
placing limits on coefficients for Lorentz and CPT violation.
The QED sectors account for more than half of the chart.
Limits in the photon sector, the largest one, 
have been reported in 77 publications to date,
with the electron, proton, and neutron sectors represented 
by 48, 34, and 27 publications, 
respectively. 
Rounding out the QED sector, 
eleven papers have presented limits on Lorentz-breaking couplings 
between matter and photons.

In the neutrino sector, 
Fig.\ \ref{sectors} shows 40 papers are referenced.
In fact,
the most recent edition of the Tables has 27 pages of neutrino-sector limits.
The quark sector is also well represented, 
with 23 publications.
In comparison, 
the lepton, electroweak, 
and gluon sectors
have fewer publications so far,
reflecting the corresponding experimental challenges 
and,
possibly, 
revealing some of the many areas where Lorentz and CPT searches  
could be very profitable.
Moving into curved spacetime, 
the gravitational sector is well represented with 35 publications to date.

\begin{figure}
\begin{center}
\includegraphics[width=180pt]{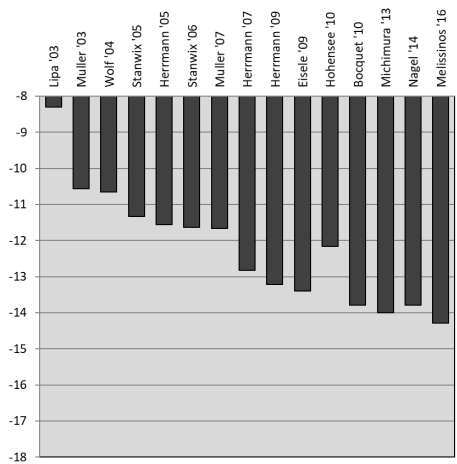}
\end{center}
\caption{Evolution of sensitivities to $({\tilde\kappa}_{o+})^{YZ}$, 
one of the 19 coefficients 
governing Lorentz-violating operators of mass dimension four 
in the photon sector. 
The scale is logarithmic, 
with the first limit placed at a few parts in $10^9$. 
The first-named author and the year of publication is listed for each result.\cite{photonexpts} 
}
\label{photon}
\end{figure}

Experimentalists have risen to the challenge 
of finding the elusive effects of Lorentz violation
by steadily improving the technology over a period of years.
As an example of this,
Fig.\ \ref{photon} 
shows the evolution in sensitivity to the 
coefficient $({\tilde\kappa}_{o+})^{YZ}$.
The first limit on this dimensionless coefficient was placed in 2003,
and up to the present it has been improved 13 times. 
The sequence of results, 
originating in labs on three continents,
spans more than a decade 
and shows the experimental reach improving by six orders of magnitude.

At present, 
the Tables show no evidence of Lorentz or CPT violation.
They also show 
that sectors with unmeasured 
or very weakly constrained coefficients exist.
The potential for finding evidence of Plank-scale physics
is of course a central motivator for the field,
and phenomenological and experimental efforts continue to grow.
As experimental technology improves,
the prospect of violations being revealed remains tantalizing.

\end{document}